\documentclass{desyproc}
\usepackage{cite}

\begin{document}
\title{Monte Carlo for the LHC
\hfill\rlap{\raisebox{1.5cm}[0pt][0pt]{\normalsize\rm
MAN/HEP/2010/17}}{}\rlap{\raisebox{1.1cm}[0pt][0pt]{\normalsize\rm
MCnet/10/15}}{}\hspace*{3cm}}


%

%

%
\author{{\slshape Michael H. Seymour}\\[1ex]
School of Physics and Astronomy, University of Manchester, Manchester,
M13 9PL, U.K.}

\contribID{xy}  
\confID{1964}
\desyproc{DESY-PROC-2010-01}
\acronym{PLHC2010}
\doi            

\maketitle

\begin{abstract}
  I review the status of the general-purpose Monte Carlo event
  generators for the LHC, with emphasis on areas of recent physics
  developments.  There has been great progress, especially in multi-jet
  simulation, but I mention some question marks that have recently
  arisen.
\end{abstract}

\section{Introduction}

There are three general-purpose Monte Carlo event generators designed
for use at the LHC, Pythia~\cite{Sjostrand:2007gs}, Herwig++\cite{Bahr:2008pv} and
Sherpa~\cite{Gleisberg:2008ta}.  The first two are built on the heritage
of their fortran predecessors~\cite{Sjostrand:2006za, Corcella:2000bw} while Sherpa has been
constructed as a new C++ project from the beginning.  Although there are
of course many differences in the details of the implementations, they
largely share a common approach to the structure of LHC events, which I
describe briefly here to set the scene and the notation.

Most of the emphasis is on the simulation of events that contain a
\emph{hard process}, although I will return to mention minimum bias
collisions later.  Since the hard interaction is generally the process
of interest it acts as the trigger around which the simulation of the
whole event is built.  In the previous generation of simulation these
were almost always $2\to2$ processes, but one of the largest areas of
development in recent years, which I will describe in detail below, has
been the inclusion of higher order corrections, both in terms of
multi-parton tree-level processes and also NLO corrections to the low
parton multiplicity processes.

The partons involved in the hard process are coloured and, just as
accelerated charges in QED radiate photons, annihilated, scattered or
produced coloured partons radiate gluons.  Now, however, unlike in QED,
since the gluons themselves are coloured, they radiate further gluons.
The hard process is therefore accompanied by an extended shower of
additional radiation, which is simulated with a \emph{parton shower\/}
algorithm.  These are formulated as a probabilistic evolution in
emission scale, from the high scale of the hard interaction downwards to
lower momentum scales.  The outgoing partons are evolved forwards to
produce a shower of accompanying radiation and the incoming partons are
evolved backwards to ask, progressively, what is the probability
distribution for radiation to accompany this parton on its way in to the
hard interaction.  Different algorithms differ in their choice of
evolution variable and can generally be split into two classes:
parton-based, as a sequence of $1\to2$ splittings with suitably-defined
(respecting the coherence of radiation from different emitters)
evolution variable and initial condition; and dipole-based, in which
colour-connected pairs of partons emit radiation as a $2\to3$ splitting,
with the colour structure taking care of the coherence condition.

As the parton shower is governed by perturbative emission probabilities
with the strong coupling evaluated at the evolution scale, it is not
valid at scales below about 1~GeV.  One therefore terminates the
evolution and invokes a non-perturbative \emph{hadronization model\/}
for the transition from a partonic to a hadronic final state.  Here
again the partonic colour structure is crucial in setting the initial
conditions for the hadronization and only models that respect this
structure, the string (Pythia) and cluster (Herwig and Sherpa) models,
are in current use.

In a hadronic collision, a partonic constituent from each hadron is
involved in the hard process and its accompanying parton shower.  The
colliding hadrons are highly Lorentz-contracted discs and in a
space-time picture completely overlap each other.  They therefore have a
high probability to have additional interactions, producing hadrons
throughout the event, in addition to those from the hadronization of the
hard process.  This is known as the \emph{underlying event\/} and is
modelled as additional independent parton--parton interactions
(multi-parton scattering models), as a soft non-perturbative interaction
of the remnants as a whole (soft underlying event models), or as a
mixture of the two.  In fact, it is essential to include a semi-hard
multi-parton interaction component to fit the HERA and Tevatron data.
Recent progress in underlying event physics has focussed on the colour
structure of the additional interactions and the colour connections
between them and the primary process and therefore simulation of the
underlying event is typically interleaved with the backward evolution of
the incoming partons.

Finally, many of the hadrons produced in the hadronization of the hard
and secondary processes are unstable resonances and their decays must be
simulated, together with other decaying particles such as the tau lepton
(decays of short-lived particles like the top, Higgs boson or SUSY
particles can be thought of as part of the hard process and are
typically simulated early in the event).  This relatively unglamorous
end of event generation has also been the subject of considerable recent
progress.

Simulation of minimum bias and diffractive collisions in which there is
no hard process is closely related to the underlying event and one
typically uses phenomenological models to describe the total rate and
its sub-division into elastic, single- and double-diffractive and
inelastic components, with the multi-jet models tuned to underlying
event data used to simulate the inelastic component.

In this talk I will give an outline of some of the most important areas
of recent physics progress.  In particular I will describe several
developments in the important area of matching parton showers with
higher order matrix elements, as well as a couple of question marks that
have recently arisen within this area.  I will describe more briefly
developments in the simulation of spin correlations, of soft
interactions and of secondary decays.  Finally I will give brief status
reports of the three general-purpose event generator projects and of
the MCnet projects for generator-independent generator validation and
tuning.

\section{Recent physics progress}

\subsection{Merging parton showers with higher order matrix elements}

Parton showers are built on approximations to the full QCD matrix
elements for multi-parton emission, expanded around the soft and
collinear limits that dominate.  They therefore perform well for the
bulk of emission.  The colour coherence of emission between different
partons in an event is crucial for this, as shown in the famous CDF
Run~1 study, \cite{Abe:1994nj}.  Three-jet events were selected with
hard two-jet kinematics, with the hardest jet being above 110~GeV and a
soft third jet only having to be above 10~GeV.  The distributions of
this third jet therefore clearly map out the radiation of the hard
$2\to2$ scattering system.  In particular, CDF's careful study showed
that coherence due to the colour connections between initial- and
final-state partons was crucial to get these distributions right.
HERWIG, which had this coherence built in, and a version of Pythia
specially modified to include it (which subsequently became the default)
were able to fit the data, while the default version of Pythia, which
included colour coherence only in final-state emission, and ISAJET,
which doesn't include it at all, were not even able to qualitatively
describe the data.  Despite the fact that this analysis is uncorrected
and more than fifteen years old, it is still an extremely important one
for Monte Carlo understanding and validation and we would dearly like to
have an update from Run~2 as well of course as looking forward to
similar analyses at the LHC.

Despite the success of modern parton shower algorithms in describing the
bulk of emissions, there are many event generator applications in which
multiple hard well-separated jets must be simulated well.  The most
obvious of these is in searches for new physics where one is often
interested in final states with many jets and where, by definition, one
designs the cuts to remove the bulk of emission so that all that remains
is the hard well-separated tail.  These regions are equally important
for the top mass measurement, QCD studies of the multi-jet regime and
many other applications.  The rate and distribution of such jets are
reasonably well described by the tree-level matrix element for the given
jet multiplicity, but parton showers are needed to describe the internal
structure of the jets and the full hadronic final state.  Moreover it is
not straightforward to merge samples with different jet multiplicities
without double-counting with subsequent emission in the shower.  Clearly
one wishes to combine the benefits of the tree-level matrix element and
parton shower approaches, and methods to do this are known as multi-jet
matching.

At the same time, there are also applications where one wishes to have
an event sample with next-to-leading order normalization, not least,
again, for new particle searches, but also for many electroweak and top
quark analyses.  Attempts to match parton showers with NLO calculations
are known as NLO matching.

Great progress has been made with both multi-jet and NLO matching over
the last five years, as I describe in the next two sections, and
practical implementations are now available for a wide variety of
processes.  Most recently, progress has been made in attempts to combine
the two approaches together, as I will also describe more briefly.

\subsubsection{Multi-jet matching}

The problem of merging tree-level matrix element samples with parton
showers for several jet multiplicities simultaneously was solved in
principle by Catani, Krauss, K\"uhn and Webber (CKKW) in
2001\cite{Catani:2001cc}.  They introduced a matching scale,
$k_{T,match}$ and showed that by using matrix elements modified by
introducing Sudakov form factors above $k_{T,match}$ and parton showers
modified by introducing appropriate phase space vetoes below
$k_{T,match}$, one could match the two in such a way that there was no
double-counting and the dependence on $k_{T,match}$ could be proved to
vanish to next-to-leading logarithmic accuracy.

However, in practical implementations, for example in the study by
Mrenna and Richardson \cite{Mrenna:2003if}, it was found that associated
distributions typically have discontinuities at $k_{T,match}$ and that
the hadron-level results were more $k_{T,match}$-dependent than the
parton-level ones.  Eventually this was explained as being due to the
CKKW method giving the right amount of radiation, as proved, but putting
some of it in the wrong place.  In particular, attributing some of it to
the wrong colour flow, affecting the initial conditions of the
hadronization phase.

This problem was solved by L\"onnblad~\cite{Lonnblad:2001iq} for the
specific case of $k_T$-ordered dipole showers and more recently for the
general parton shower case by the Sherpa~\cite{Hoeche:2009rj} and
Herwig++\cite{Hamilton:2009ne} collaborations following an idea
originally proposed by Nason~\cite{Nason:2004rx} as part of the POWHEG
approach described below.  The idea is that one should run the parton
shower from the lowest multiplicity configuration and to forcibly insert
emissions corresponding to the exact kinematics generated by the matrix
element event into the appropriate point in the ordering of the shower.
The effect is to generate \emph{truncated showers\/} from the internal
lines of the matrix element event, as well as the external lines, and to
properly populate the whole of phase space with soft radiation with the
correct colour connections.

Results of this modified CKKW method have been compared with data from
the Tevatron, for example the CDF W+jets~\cite{Aaltonen:2007ip} and
Z+jets~\cite{:2007cp} data, in \cite{Hoeche:2009rj,Hamilton:2009ne}.
Rates and distributions of events with up to four jets are well
described and the residual matching scale dependence is shown to be very
small, with a corresponding uncertainty in the total cross section of
only 3\%.

\subsubsection{NLO matching}

In a conventional Monte Carlo implementation of a next-to-leading order
calculation, events in the real emission phase space have arbitrarily
large positive weights, which are cancelled to give a finite cross
section contribution by counter-events that have equally large negative
weights but live in the phase space of the Born process.  The result is
finite for any infrared safe observable, but the procedure is not
suitable for implementation into a parton shower, hadronization and
detector simulation framework, since any arbitrarily small differences
in the subsequent final state of the event and counter-event would spoil
the cancellation.

Frixione and Webber showed in 2002\cite{Frixione:2002ik} that this
problem could be solved to give Monte Carlo events with finite weight
distribution, essentially by using an analytical expansion of the parton
shower emission probability as the subtraction counter-event term.  The
result is a set of hard + either 1-jet or 0-jet events to be showered,
such that there is no double-counting between the showered 0-jet and
generated 1-jet events.  Although the weight distributions are finite,
they are not positive definite and one typically generates `almost
unweighted' events, i.e.\ with equal absolute values of weights, but
typically around 10\% of them negative.  This is not a problem of
principle, but can be inconvenient for some applications.  A more
serious problem is the fact that the analytical subtractions have to be
calculated for the particular parton shower with which it will be used
and, thus far, this MC@NLO method is available for a wide range of
processes~\cite{Frixione:2008ym} only for use with the original HERWIG
program.  With a first implementation for PYTHIA reported in
Ref.~\cite{Torrielli:2010aw}, a full version for both PYTHIA and
Herwig++ is expected to appear soon.

However, a potentially more serious problem with the MC@NLO approach was
noticed in Ref.~\cite{Mangano:2006rw} and explored in more detail in
Ref.~\cite{Alioli:2008tz}.  It is that MC@NLO distributions can inherit
deficiencies in the underlying shower algorithm.  This is most evident
in the jet rapidity distributions in which the PYTHIA and especially
HERWIG algorithms produce insufficient hard central jets.  Although the
MC@NLO algorithm corrects this distribution analytically to leading
order, all higher orders are directly inherited from the shower.  The
result, especially in gluon-initiated processes such as $gg\to H$, can
be rapidity distributions with significant unphysical `notches' in them,
see for example Fig.~9 of~\cite{Alioli:2008tz}.

A second issue with the MC@NLO approach is that it is guaranteed to
exactly reproduce the LO ``+1-jet'' cross section at high enough
transverse momentum.  This may sound like a good feature, but it turns
out that processes for which the $K$~factor is significant, so for which
one definitely wants to use a NLO matching approach, the $K$~factor for
the ``+1-jet'' process is also large, so that the MC@NLO result is
significantly below the NLO result for the high-$p_t$ distribution.  The
most extreme case is again $gg\to H$ where the difference is around a
factor of two.

While in both of these cases, MC@NLO is formally correct to the order at
which it is defined (NLO for the normalization and LO for the $p_t$
distribution), phenomenologically one might prefer a solution that does
not suffer from these effects.  This is provided by the POWHEG method
proposed by Nason~\cite{Nason:2004rx} in 2004.  It has the advantages that
it provides only positive-weight events, that the distribution of
hardest emission is entirely determined by the hard matrix element,
without inheriting features from the parton shower, and that the entire
distribution receives the $K$~factor so, to the extent that the
$K$~factors of the inclusive and high-$p_t$ processes are similar, the
latter is well described.  Finally it is independent of shower
algorithm and can be used with any parton shower algorithm that is
capable of producing the truncated showers discussed earlier.

The POWHEG method is implemented as a standalone program, also called
POWHEG, for an increasingly wide range of processes~\cite{Alioli:2008tz,
  Frixione:2007vw, Alioli:2008gx, Alioli:2009je, Nason:2009ai,
  Alioli:2010xd, Re:2010jg}.  It has also become the method of choice
for NLO matching in the Herwig++ program, which also now comes with
built-in POWHEG implementations for Drell-Yan
production~\cite{Hamilton:2008pd}, Higgs production~\cite{Hamilton:2009za}
and $e^+e^-$ processes~\cite{LatundeDada:2008bv}, with vector boson
fusion, deep inelastic scattering and vector boson pair production
including anomalous triple gauge couplings in progress.  The deep
inelastic scattering implementation in particular allows Herwig++ to
describe the energy flow data from the HERA experiments over a wide
range of $x$ and $Q^2$ for the first time.

\subsubsection{Towards NLO multi-jet matching}

Given the success of multi-jet and NLO matching schemes, it is natural
to ask whether they can be combined to produce a multi-jet sample in
which each of the jet multiplicities is correct to NLO.  Ideas towards
this ambitious goal have been described in Ref.~\cite{Nagy:2005aa}.  In
Ref.~\cite{Lavesson:2008ah} the first concrete implementation appeared,
only for the case of $e^+e^-$ annihilation.  The extension to hadron
collisions is considerably more complicated and is yet to appear as a
working implementation.

Hamilton and Nason~\cite{Hamilton:2010wh} took a more pragmatic approach.
Motivated by the large body of validated multi-jet and NLO matching
implementations in use, they examined whether it is possible to combine
the POWHEG and CKKW approaches to provide a sample of multi-jet events,
each calculated using the tree-level multi-parton matrix element
combined with the full NLO correction for the Born configuration onto
which it is mapped.  They succeeded in this and studied implementations
for vector boson and top pair production in hadron collisions.

With the progress made in these approaches it seems hopeful that a
working NLO multi-jet matching algorithm could be achieved in the near
future.  It is clear that this would be a major step forward in our
ability to simulate LHC final states.

\subsubsection{High energy jets}

I previously said that parton shower algorithms do well for the bulk of
emission, but with the large step up in energy to the LHC, and the
consequent opening up of phase space, we should constantly question this
statement and check that we are sure.  In this and the next section I
mention two recent calculations that raise small question marks over our
readiness.

Andersen and collaborators~\cite{Andersen:2008ue, Andersen:2009nu} have
developed a new approach to calculating multi-jet final states, which
they call the high energy jet (HEJ) approach.  It relies on
approximating the all-order QCD matrix elements in a different limit to
parton showers, namely the limit of fixed momentum transfer with
available scattering energy going to infinity, the limit in which many
jets have similar transverse momenta and large rapidity intervals.  They
show that in this limit, scattering amplitudes factorize into
helicity-dependent local terms, coupled by $t$-channel propagators,
which can be constructed in a modular way to arbitrary order.  They have
working
implementations for pure jet processes, W/Z plus jets and Higgs plus
jets and have made a thorough phenomenological analysis.  As an example,
see Fig.~69 of Ref.~\cite{Jeppe3} in which the results are compared with
the Sherpa shower with CKKW matching and the NLO calculation in MCFM,
for the accompanying jet multiplicity of Higgs plus at least two jet
events as a function of the rapidity separation between the two leading
jets.  For small separations all three calculations agree, but by
$\Delta y=4$, a typical cut used to separate this gluon fusion process
from the vector boson fusion process, the HEJ approach is significantly
above the other two, predicting an average number of additional jets of
order~1.  By $\Delta y=6$, still within the typical region of a VBF
analysis, HEJ predicts twice as many additional jets as either CKKW or
NLO.

The HEJ code exists as a working Monte Carlo and work is in progress to
match it properly with parton showers.  It will be extremely interesting
to see it further used to validate the existing parton shower and
matching algorithms and to see whether it can be developed to become a
fully-fledged alternative to multi-jet matching (here one could mention
that it is much faster than calculating the full multi-jet matrix
elements for high jet multiplicities).

\subsubsection{Giant \boldmath$K$~factors}

It has been known for some time that certain observables suffer from
anomalously large $K$~factors.  In a recent study~\cite{Rubin:2010xp},
Rubin, Salam and Sapeta considered this in more detail, isolated the
origin of these giant $K$~factors and showed how to calculate the next
higher order in such cases to stabilize the perturbative series.  In
this section I consider the connection with parton shower algorithms.

The archetypal process in which they study this is Z+jets at high $p_t$
(see Fig.~1 of Ref.~\cite{Rubin:2010xp}).  In the Z $p_t$ spectrum the
$K$~factor is roughly constant at about 1.5 and just about consistent
with the scale variation: the LO and NLO bands just touch.  However,
turning to the $p_t$ distribution of the leading jet, which is
equivalent at leading order, they found a $K$~factor that grows linearly
with $p_t$ from about 2 at 250~GeV to more than 5 at 1~TeV and that is
in no way represented by the scale variation.  Finally, they considered
an observable that is important for search physics, the total scalar
transverse momentum of all jets that accompany the Z, $H_T$, again
equivalent at leading order.  They found that the $K$~factor grows
exponentially with $H_T$, from about 10 at 500~GeV to 1000 at 2.5~TeV.

They argued that the large $K$~factor in the leading jet $p_t$
distribution is due to a new kinematic regime opening up, namely the
possibility that two hard jets could be produced, accompanied by a
relatively soft Z boson.  It has long been known that electroweak
corrections to high $p_t$ jet production are large and negative owing to
an electroweak Sudakov form factor with leading order term
$\sim-\alpha_W\log^2p_t/M_z$.  Its counterpart is a real correction to
dijet production $\sim+\alpha_W\log^2p_t/M_z$ due to the emission of a Z
boson.  This can equivalently be seen, in our case, as a real correction
$\sim+\alpha_s\log^2p_t/M_z$ to the Z+jet process.  One can easily check
that this dependence is roughly linear over the $p_t$ range considered
and of the same order of magnitude as the NLO correction actually seen.
Finally, this understanding also allows an understanding of the huge
$K$~factor seen in the $H_T$ distribution: the events with two
high-$p_t$ jets and a low-$p_t$ Z, which occur at about the same rate as
Z+one-jet events with the same jet $p_t$, contribute to a value of $H_T$
a factor of two higher than $p_t$.  Since the underlying LO $p_t$
distribution is falling so rapidly, this factor of 2 increase in the
value of the observable corresponds to a huge, exponential, increase in
the value of the cross section at a given value of the observable.

In Ref.~\cite{Rubin:2010xp} an ingenious method was proposed to
calculate higher order corrections to such processes.  The main point is
that a unitarity-type argument is used to estimate uncalculated loop
corrections from calculated tree-level corrections at the same order.
This process has similarities with the CKKW idea and I believe this
connection could be explored further, but here I confine myself to
drawing conclusions for Monte Carlo event generators.

The phase space region responsible for these large corrections
corresponds to $2\to2$ QCD scattering events in which a Z boson is
radiated from an incoming or outgoing quark or antiquark.  Such W and Z
parton shower radiation is not implemented in any of the general-purpose
generators, despite having been identified as important in
Ref.~\cite{thesis}.  Although this effect should be reproduced by the
CKKW method, for a smooth matching, for systematic studies and for
processes in which one does not have a CKKW implementation, one should
include as much of the relevant physics in the shower as possible.  It
is clear that as we enter the LHC era the need to include electroweak
boson radiation is more urgent.

\subsection{Spin correlations}

Spin correlations play an extremely important role in many event
generator applications.  For example in some searches for BSM physics
one is interested in cascade decays in which the angular distributions
are crucial for determining the spins of the decaying
particles~\cite{Athanasiou:2006ef}.  Sherpa~\cite{Gleisberg:2008ta} and
Herwig++\cite{Bahr:2008pv} both have spin correlations built in in a
flexible way.  The classic example, on which both have been extensively
validated, is in tau physics.  For example in Higgs decays to
$\tau^+\tau^-$ with both taus decaying to a single pion, one can
determine whether the Higgs is a scalar or a pseudoscalar from the
azimuthal correlation between the two decay planes.  Both programs have
been shown to reproduce the analytical result~\cite{Kramer:1993jn} well.

\subsection{Underlying events/minimum bias/diffraction}

These have been a traditional strength of Pythia, with highly
developed multi-parton interaction and soft diffraction models.  A
recent development has been the inclusion of a hard diffractive
component~\cite{Navin:2010kk} into Pythia~8 along similar lines to the older
standalone program Pompyt.

Herwig++ and Sherpa are also catching up in this area, with Herwig++
having a multi-parton interaction model developed from the Jimmy
program, but with the addition of soft parton--parton scattering
allowing simulation of minimum bias collisions for the first
time~\cite{Bahr:2008dy, Bahr:2008wk, Bahr:2009ek}.  A forthcoming version, with the further addition of
colour correlation effects between the scatters, appears to be able to
describe the ATLAS data~\cite{Aad:2010rd}, with detailed tuning currently in
progress.  Sherpa also has a new minimum bias model~\cite{Gleisberg:2008ta}
which looks promising.

\subsection{Secondary decays}

Both Herwig++~\cite{Bahr:2008pv, Grellscheid:2007tt} and Sherpa~\cite{Gleisberg:2008ta} have
implemented extensive secondary decay models, with detailed matrix
elements for a wide variety of final states and interference with many
intermediate resonances, and spin correlations between decays.
Moreover, both include QED corrections in the YFS
scheme~\cite{Hamilton:2006xz, Schonherr:2008av}.  The aim is to have at least
as good a description as EVTGEN, TAUOLA and PHOTOS in all cases and
thereby dispense with the need for such external packages and their
results have been extensively validated against these programs.

\section{Status reports}

I finish my talk with very brief status reports of the main event
generator projects.  More detailed and up-to-date information can always
be obtained from the web sites listed below.

\subsection{Pythia}

The fortran Pythia~6 program, which has been the workhorse of particle
physics for some 25 years is still supported but is not being actively
developed.  All new physics developments go into the Pythia~8 program.
Its core is ready and tuned, with a much more flexible structure to
allow for the extensive physics model development that is now ongoing.
Some features of Pythia~6 are definitely dropped, for example the old
virtuality-ordered showers, and many new features added, for example
hard scattering in diffraction, a significantly improved underlying
event treatment and wide range of new BSMs.

\noindent\texttt{http://projects.hepforge.org/pythia}

\subsection{Herwig}

The current version of the fortran HERWIG program has been effectively
frozen for three years, but a bug fix release will appear this summer.
All development is now transferred to Herwig++, which has many physics
improvements, including improved angular-ordered parton showers, with
facilities built in to match with multi-jet or POWHEG hard processes, a
slightly improved implementation of the cluster hadronization model, the
improvements to soft modelling that allow minimum bias to be simulated
for the first time and a very flexible framework for implementing new
physics models.  One advantage over HERWIG is the fact that each version
is released with a globally-fitted parameter set.  A new version release
is expected this summer and should fully replace HERWIG.

\noindent\texttt{http://projects.hepforge.org/herwig}

\subsection{Sherpa}

Unlike the previous two generators, Sherpa was designed as a new
generator in C++ from the start.  In order to get started it had
interfaces to external packages for some components, but by now it is a
fully-fledged standalone generator.  The emphasis is on multi-jet final
states, with two different automated high-multiplicity matrix element
generators, an automated subtraction algorithm for NLO calculations, a
$k_t$-ordered dipole shower and built-in CKKW matching.  It also has a
multi-parton interaction model and a new cluster hadronization model.

\noindent\texttt{http://projects.hepforge.org/sherpa}

\subsection{Tuning and validation}

Within MCnet there are two other important Monte Carlo projects,
Rivet~\cite{Buckley:2010ar}, a generator- and experiment-independent framework
for validation of generators against experiment, and
Professor~\cite{Buckley:2009bj}, a generator-independent semi-automated
parameter tuning tool.  With Rivet, the Tevatron experiments are
starting to develop a culture, which was prevalent with its predecessor
HZTOOL with the HERA experiments, that all important analyses get
immediately implemented to ensure that the full details of the analysis
get preserved for posterity and the data can be compared to theory
calculations and models on an exactly like-for-like basis for years to
come.  It is essential that this culture continue at the LHC, to ensure
that its data gets fully preserved and exploited.  As well as its
important function in tuning event generator parameters, Professor
provides a set of tools for visualizing the data and seeing in real time
how it responds to particular combinations of parameter settings.

Both of these tools are being incorporated into the LHC experiments'
software frameworks, to ensure that models tuned to the new data
continue to describe the existing data, the first time any experiments
have done this in such detail.

\noindent\texttt{http://projects.hepforge.org/rivet}

\noindent\texttt{http://projects.hepforge.org/professor}

\section{Summary}

Modern Monte Carlo event generators are highly sophisticated
implementations of QCD calculations.  They are reliable for a wide
variety of observables over a wide range of energy scales and the
model-dependent parts widely validated.  But the LHC is a truly huge
step into the unknown, requiring extensive tuning of soft models and
validation of hard evolution.  There has been a great deal of progress
in describing hard emission more accurately, but, as I have shown, a few
small areas where more work is needed.

\section*{Acknowledgments}

I am grateful to all the other Monte Carlo authors and members of the
MCnet network for discussions and input.  Thanks especially to Peter
Richardson, Peter Skands, Stefan H\"oche, Frank Siegert, Jeppe Andersen,
Gavin Salam and Mathieu Rubin for contributing slides and plots to the
talk.  This work was supported in part by the European Union FP6 Marie
Curie Research Training Network MCnet (contract MRTN-CT-2006-035606).
 

\begin{footnotesize}

\end{footnotesize}



\begin{thebibliography}{99}


\bibitem{Sjostrand:2007gs}
  T.~Sj\"ostrand, S.~Mrenna and P.~Z.~Skands,
  Comput.\ Phys.\ Commun.\  {\bf 178} (2008) 852
  [arXiv:0710.3820 [hep-ph]].
\bibitem{Bahr:2008pv}
  M.~B\"ahr {\it et al.},
  Eur.\ Phys.\ J.\  C {\bf 58} (2008) 639
  [arXiv:0803.0883 [hep-ph]].
\bibitem{Gleisberg:2008ta}
  T.~Gleisberg, S.~H\"oche, F.~Krauss, M.~Sch\"onherr, S.~Schumann, F.~Siegert and J.~Winter,
  JHEP {\bf 0902} (2009) 007
  [arXiv:0811.4622 [hep-ph]].
\bibitem{Sjostrand:2006za}
  T.~Sj\"ostrand, S.~Mrenna and P.~Z.~Skands,
  JHEP {\bf 0605} (2006) 026
  [arXiv:hep-ph/0603175].
\bibitem{Corcella:2000bw}
  G.~Corcella {\it et al.},
  JHEP {\bf 0101} (2001) 010
  [arXiv:hep-ph/0011363].
\bibitem{Abe:1994nj}
  F.~Abe {\it et al.}  [CDF Collaboration],
  Phys.\ Rev.\  D {\bf 50} (1994) 5562.
\bibitem{Catani:2001cc}
  S.~Catani, F.~Krauss, R.~K\"uhn and B.~R.~Webber,
  JHEP {\bf 0111} (2001) 063
  [arXiv:hep-ph/0109231].
\bibitem{Mrenna:2003if}
  S.~Mrenna and P.~Richardson,
  JHEP {\bf 0405} (2004) 040
  [arXiv:hep-ph/0312274].
\bibitem{Lonnblad:2001iq}
  L.~L\"onnblad,
  JHEP {\bf 0205} (2002) 046
  [arXiv:hep-ph/0112284].
\bibitem{Hoeche:2009rj}
  S.~H\"oche, F.~Krauss, S.~Schumann and F.~Siegert,
  JHEP {\bf 0905} (2009) 053
  [arXiv:0903.1219 [hep-ph]].
\bibitem{Hamilton:2009ne}
  K.~Hamilton, P.~Richardson and J.~Tully,
  JHEP {\bf 0911} (2009) 038
  [arXiv:0905.3072 [hep-ph]].
\bibitem{Nason:2004rx}
  P.~Nason,
  JHEP {\bf 0411} (2004) 040
  [arXiv:hep-ph/0409146].
\bibitem{Aaltonen:2007ip}
  T.~Aaltonen {\it et al.}  [CDF Collaboration],
  Phys.\ Rev.\  D {\bf 77} (2008) 011108
  [arXiv:0711.4044 [hep-ex]].
\bibitem{:2007cp}
  T.~Aaltonen {\it et al.}  [CDF Collaboration],
  Phys.\ Rev.\ Lett.\  {\bf 100} (2008) 102001
  [arXiv:0711.3717 [hep-ex]].
\bibitem{Frixione:2002ik}
  S.~Frixione and B.~R.~Webber,
  JHEP {\bf 0206} (2002) 029
  [arXiv:hep-ph/0204244].
\bibitem{Frixione:2008ym}
  S.~Frixione and B.~R.~Webber,
  arXiv:0812.0770 [hep-ph].
\bibitem{Torrielli:2010aw}
  P.~Torrielli and S.~Frixione,
  JHEP {\bf 1004}, 110 (2010)
  [arXiv:1002.4293 [hep-ph]].
\bibitem{Mangano:2006rw}
  M.~L.~Mangano, M.~Moretti, F.~Piccinini and M.~Treccani,
  JHEP {\bf 0701} (2007) 013
  [arXiv:hep-ph/0611129].
\bibitem{Alioli:2008tz}
  S.~Alioli, P.~Nason, C.~Oleari and E.~Re,
  JHEP {\bf 0904}, 002 (2009)
  [arXiv:0812.0578 [hep-ph]].
\bibitem{Frixione:2007vw}
  S.~Frixione, P.~Nason and C.~Oleari,
  JHEP {\bf 0711}, 070 (2007)
  [arXiv:0709.2092 [hep-ph]].
\bibitem{Alioli:2008gx}
  S.~Alioli, P.~Nason, C.~Oleari and E.~Re,
  JHEP {\bf 0807}, 060 (2008)
  [arXiv:0805.4802 [hep-ph]].
\bibitem{Alioli:2009je}
  S.~Alioli, P.~Nason, C.~Oleari and E.~Re,
  JHEP {\bf 0909}, 111 (2009)
  [Erratum-ibid.\  {\bf 1002}, 011 (2010)]
  [arXiv:0907.4076 [hep-ph]].
\bibitem{Nason:2009ai}
  P.~Nason and C.~Oleari,
  JHEP {\bf 1002}, 037 (2010)
  [arXiv:0911.5299 [hep-ph]].
\bibitem{Alioli:2010xd}
  S.~Alioli, P.~Nason, C.~Oleari and E.~Re,
  JHEP {\bf 1006}, 043 (2010)
  [arXiv:1002.2581 [hep-ph]].
\bibitem{Re:2010jg}
  E.~Re,
  arXiv:1007.0498 [hep-ph].
\bibitem{Hamilton:2008pd}
  K.~Hamilton, P.~Richardson and J.~Tully,
  JHEP {\bf 0810} (2008) 015
  [arXiv:0806.0290 [hep-ph]].
\bibitem{Hamilton:2009za}
  K.~Hamilton, P.~Richardson and J.~Tully,
  JHEP {\bf 0904} (2009) 116
  [arXiv:0903.4345 [hep-ph]].
\bibitem{LatundeDada:2008bv}
  O.~Latunde-Dada,
  Eur.\ Phys.\ J.\  C {\bf 58}, 543 (2008)
  [arXiv:0806.4560 [hep-ph]].
\bibitem{Nagy:2005aa}
  Z.~Nagy and D.~E.~Soper,
  JHEP {\bf 0510} (2005) 024
  [arXiv:hep-ph/0503053].
\bibitem{Lavesson:2008ah}
  N.~Lavesson and L.~L\"onnblad,
  JHEP {\bf 0812} (2008) 070
  [arXiv:0811.2912 [hep-ph]].
\bibitem{Hamilton:2010wh}
  K.~Hamilton and P.~Nason,
  JHEP {\bf 1006} (2010) 039
  [arXiv:1004.1764 [hep-ph]].
\bibitem{Andersen:2008ue}
  J.~R.~Andersen and C.~D.~White,
  Phys.\ Rev.\  D {\bf 78} (2008) 051501
  [arXiv:0802.2858 [hep-ph]].
\bibitem{Andersen:2009nu}
  J.~R.~Andersen and J.~M.~Smillie,
  JHEP {\bf 1001} (2010) 039
  [arXiv:0908.2786 [hep-ph]].
\bibitem{Jeppe3}
  J.~R.~Andersen, J.~Campbell and S.~H\"oche, in \cite{Binoth:2010ra},
  pp.~130--132.
\bibitem{Binoth:2010ra}
  T.~Binoth {\it et al.}  [SM and NLO Multileg Working Group],
  arXiv:1003.1241 [hep-ph].
\bibitem{Rubin:2010xp}
  M.~Rubin, G.~P.~Salam and S.~Sapeta,
  arXiv:1006.2144 [hep-ph].
\bibitem{thesis}
  M.~H.~Seymour,
  University of Cambridge PhD thesis, 1992.
\bibitem{Athanasiou:2006ef}
  C.~Athanasiou, C.~G.~Lester, J.~M.~Smillie and B.~R.~Webber,
  JHEP {\bf 0608} (2006) 055
  [arXiv:hep-ph/0605286].
\bibitem{Kramer:1993jn}
  M.~Kr\"amer, J.~H.~Kuhn, M.~L.~Stong and P.~M.~Zerwas,
  Z.\ Phys.\  C {\bf 64} (1994) 21
  [arXiv:hep-ph/9404280].
\bibitem{Navin:2010kk}
  S.~Navin,
  arXiv:1005.3894 [hep-ph].
\bibitem{Bahr:2008dy}
  M.~B\"ahr, S.~Gieseke and M.~H.~Seymour,
  JHEP {\bf 0807} (2008) 076
  [arXiv:0803.3633 [hep-ph]].
\bibitem{Bahr:2008wk}
  M.~B\"ahr, J.~M.~Butterworth and M.~H.~Seymour,
  JHEP {\bf 0901} (2009) 065
  [arXiv:0806.2949 [hep-ph]].
\bibitem{Bahr:2009ek}
  M.~B\"ahr, J.~M.~Butterworth, S.~Gieseke and M.~H.~Seymour,
  arXiv:0905.4671 [hep-ph].
\bibitem{Aad:2010rd}
  G.~Aad {\it et al.}  [ATLAS Collaboration],
  Phys.\ Lett.\  B {\bf 688} (2010) 21
  [arXiv:1003.3124 [hep-ex]].
\bibitem{Grellscheid:2007tt}
  D.~Grellscheid and P.~Richardson,
  arXiv:0710.1951 [hep-ph].
\bibitem{Schonherr:2008av}
 M.~Sch\"onherr and F.~Krauss,
 JHEP {\bf 0812} (2008) 018
 [arXiv:0810.5071 [hep-ph]].
\bibitem{Hamilton:2006xz}
  K.~Hamilton and P.~Richardson,
  JHEP {\bf 0607} (2006) 010
  [arXiv:hep-ph/0603034].
\bibitem{Buckley:2010ar}
  A.~Buckley {\it et al.},
  arXiv:1003.0694 [hep-ph].
\bibitem{Buckley:2009bj}
  A.~Buckley, H.~Hoeth, H.~Lacker, H.~Schulz and J.~E.~von Seggern,
  Eur.\ Phys.\ J.\  C {\bf 65} (2010) 331
  [arXiv:0907.2973 [hep-ph]].
\end{thebibliography}
\end{document}